\documentstyle[twocolumn,aps]{revtex} 

\draft 
\begin{document} 
 
 
\title{Expansion of the Vortex Cores
in YBa$_2$Cu$_3$O$_{6.95}$ at Low Magnetic Fields} 
 
\author{ J.~E.~Sonier\thanks{Current address: Los Alamos National Laboratory,
Los Alamos, New Mexico 87545}$^{,1}$, R.~F.~Kiefl$^1$, J.~H.~Brewer$^1$, D.A.~Bonn$^1$, 
S.R.~Dunsiger$^1$, W.N.~Hardy$^1$, R.~Liang$^1$, R.~I.~Miller$^1$,
D.~R.~Noakes$^2$ and C.~E.~Stronach$^2$}

\address{$^1$TRIUMF, Canadian Institute for Advanced Research 
 and Department of Physics and Astronomy, University of British Columbia, 
 Vancouver, British Columbia, Canada V6T 1Z1 } 
\address{$^2$Department of Physics, Virginia State University, 
Petersburg, Virginia 23806}

\date{June 1, 1998} 
\date{ \rule{2.5in}{0pt} } 
 
\maketitle 
\begin{abstract} \noindent 
Muon spin rotation ($\mu$SR) spectroscopy has been used to measure
the effective size $r_0$ of the vortex cores in optimally doped 
YBa$_2$Cu$_3$O$_{6.95}$ as a function of
temperature $T$ and magnetic field $H$ deep in the superconducting state. 
While $r_0$ at $H \! = \! 2$~T is close to 20~\AA~ and consistent with
that measured by STM at 6~T, we find a striking increase in $r_0$
at lower magnetic fields, where it approaches an extraordinarily large
value of about 100~\AA. This suggests that the
average value of the superconducting coherence
length $\xi_{ab}$ in cuprate superconductors may be much larger
than previously thought at low magnetic fields in the vortex state.
\end{abstract} 
\newpage 

The vortex core in a type-II superconductor is a region in which
the superconducting order parameter $\psi {(\bf r)}$ is strongly suppressed.
The size of the vortex core is therefore closely related to the coherence
length $\xi$---which is the smallest length over which $\psi {(\bf r)}$
can change appreciably. Some time ago Caroli {\it et al.} \cite{Caroli:64}
predicted that a discrete spectrum of quasiparticle
excitations existed within a
radius $\xi$ of the vortex axis. An
STM experiment on the conventional 
type-II superconductor NbSe$_2$ by Hess {\it et al.} \cite{Hess:89} 
confirmed the existence of these localized states in the vortex core. 
Since then both $\mu$SR \cite{Sonier:97}
and STM \cite{Hartmann:93} measurements have shown that the 
vortex core size $r_0$ in NbSe$_2$ decreases with increasing
magnetic field---in a manner which scales with the increased strength of the
vortex-vortex interactions. These same techniques
\cite{Sonier:97,Volodin:97,Miller:98} have also shown that $r_0$ shrinks with
decreasing temperature, as predicted
by the so-called ``Kramer-Pesch effect'' \cite{Kramer:74}.

The electronic structure of the vortex cores in the high-$T_c$ superconductors
(HTSs)
is less certain. Experiments performed on YBa$_2$Cu$_3$O$_{7-\delta}$ (YBCO)
\cite{Maggio:95,Karrai:92} and Nd$_{1.85}$Ce$_{0.15}$CuO$_{4-\delta}$ (NCCO)
\cite{Jiang:95} seem to support the existence of a
few bound quasiparticle states in the vortex core.
On the other hand, no evidence of such localized states was found
in a recent STM study of the vortex core in under and overdoped 
Bi$_2$Sr$_2$CaCu$_2$O$_{8+\delta}$ (BSSCO) \cite{Renner:98}. Instead the
tunneling spectra in the vortex core resembled the
pseudogap which forms at the Fermi surface in the normal state
of this material.
It should be noted that the vortex lattice studied in Ref.~\cite{Renner:98}
was either melted or consisted of randomly pinned pancake vortices.

It is now widely accepted that the order parameter in the hole-doped
HTSs (which excludes NCCO) have a $d_{x^2-y^2}$-wave symmetry.
Several authors
have pointed out \cite{Wang:95,Ichioka:96} 
that bound quasiparticle states are unlikely to exist in a vortex of
a $d_{x^2-y^2}$-wave superconductor because of the nodes which are
present in the
energy gap function $\Delta_{\hat{k}} \! = \! \Delta_0 (\hat{k}_x^2 \!
- \! \hat{k}_y^2)$ along the 
directions $| \hat{k}_x | \! = \! | \hat{k}_y |$.
One way of explaining experiments on HTSs which support localized
states in the vortex core, is to introduce 
additional components into $\psi({\bf r})$
({\it e.g.} a $d_{xy}$-wave component \cite{Franz:98}). 
However, there is currently no
direct evidence for an order parameter of mixed symmetry in the 
bulk of HTSs.

In a previous study \cite{Sonier:97b}, we
determined the size of the 
vortex cores in the underdoped compound YBa$_2$Cu$_3$O$_{6.60}$.
The electronic structure of the cores in underdoped YBCO has
yet to be investigated with STM---although one may anticipate a local
density of states resembling the normal state pseudogap, as in
underdoped BSCCO \cite{Renner:98}. In Ref.~\cite{Sonier:97b}
$r_0$ was found to change as a function of $T$ and $H$ in a manner
similar to that observed
in NbSe$_2$, but with a considerably weaker temperature
dependence. 
Hayashi {\it et al.} \cite{Hayashi:98} 
recently suggested that this
may indicate that the quantum limit is established at 
a much higher temperature in YBa$_2$Cu$_3$O$_{6.60}$ because
of the small value of $\xi$ relative to that in NbSe$_2$.
Generally speaking, $\xi$ is considered to be ``short''
in the HTSs ({\it e.g.} $< \! 20$~\AA) \cite{Hc2}.
This is one of the primary features which distinguishes them
from conventional superconductors.
In YBa$_2$Cu$_3$O$_{6.60}$, $r_0$ was found to be as large as 80~\AA~
at low magnetic fields, which suggests that $\xi$ is in excess of 20~\AA.
However, this is a phase with a $T_c$ of 60~K, so $\xi$ is 
expected to be somewhat larger than in a higher $T_c$ sample.
Thus, it is of great interest to check the
size and field dependence of $r_0$ in optimally doped 
YBa$_2$Cu$_3$O$_{6.95}$, with the maximum $T_c$ of 93~K. 
Unlike in underdoped YBCO, the formation of the pseudogap
in YBa$_2$Cu$_3$O$_{6.95}$ appears to coincide with the superconducting
gap at $T_c$. The STM study of the vortex core by
Maggio-Aprile {\it et al.}
\cite{Maggio:95} in near optimally-doped YBCO, clearly
shows two peaks within a gap like structure---indicating the
presence of bound quasiparticle states.

In this Letter we present $\mu$SR measurements of the effective vortex core
size in YBa$_2$Cu$_3$O$_{6.95}$ as a function of 
temperature and magnetic field applied 
along the ${\bf\hat{c}}$ axis of the crystals.
Our measurements of $r_0$ are shown to be
consistent with the STM study 
of Ref.~\cite{Maggio:95} at $H \! = \! 6$~T. Surprisingly,
we find that at low fields $r_0$ increases to a comparatively large value
of 100~\AA. A simple interpretation
of this result would be that the length scale over which the
order parameter changes in the region of the vortex core
({\it i.e.} the definition of the coherence length as
pertaining to the vortex core) is considerably larger
than the nominal and accepted value of about 20~\AA~
measured in high magnetic fields. 

We report here measurements of $r_0$ in three 
different samples of YBa$_2$Cu$_3$O$_{6.95}$.
The magnetic field distributions in two of these samples 
were previously recorded \cite{Sonier:97c}.
The first (TW1) was a mosaic of three crystals whereas the
other (TW2) was a single crystal. All of these crystals contained 
twin boundaries and had transition temperatures of 93.2(0.25)~K.
We also report here measurements taken in
TW2 after removing the twin boundaries.
Detwinning was achieved by applying uniaxial stress to the crystal
with the sample 
heated to no more than 250~$^\circ$C in an oxygen atmosphere. 
Subsequent to the  mechanical detwinning process, 
the crystal was reannealed to set the oxygen doping level.
The $\mu$SR experiments were performed on the M15 and M20 surface beamlines
at TRIUMF. Our experimental setup is described elsewhere \cite{Kiefl:94}.

Although $\mu$SR does not directly probe the electronic
structure of the vortex cores, it does sample the distribution
of local magnetic fields in the vicinity of the cores.
The spin of an implanted muon precesses at a frequency which
is directly proportional to the local magnetic field at the muon site. 
Since the local magnetic field rises to a maximum in the vortex cores where
superconductivity is destroyed, $r_0$ is directly related to the high-field
tail in the measured field distribution. The size of the core is not
strictly defined however, because 
there is no sharp discontinuity in spatial quantities
between a normal vortex core and the surrounding
superconducting material.
Here, as in our
previous work \cite{Sonier:97,Sonier:97b}, we define $r_0$ to be the
radius about the vortex axis at which the supercurrent 
density $J_s({\bf r})$ reaches its maximum value. 
This feature allows us to accurately monitor changes in the effective
size of the vortex cores.
The supercurrent density $J_s ({\bf r})$ is obtained from the field
profile ${\bf B} ({\bf r})$ through the Maxwell relation
$J_s ({\bf r}) \! = \! | \mbox{\boldmath $\nabla$} \! \times \!
{\bf B} ({\bf r}) |$. In fitting the measured muon spin precession signal,
some modelling of ${\bf B} ({\bf r})$ is required.
However, to appreciate the accuracy of the present study
it is important to realize that the $J_s ({\bf r})$
profile does not depend on the validity of the model assumed, since
it is essentially the same for any
function ${\bf B} ({\bf r})$ which fits the data well.

As in Refs.~\cite{Sonier:97,Sonier:97b}, the 
local field due to the vortex lattice at any point in the $\hat{a}$-%
$\hat{b}$ plane was 
modelled with a theoretical field distribution generated from a
Ginzburg-Landau (GL) model \cite{Yaouanc:97}
\begin{eqnarray} 
\label{eq:Br} 
 B(\mbox{\boldmath r}) & = & B_0 (1-b^4)\sum_{ {\bf G} } 
 { e^{-i {\bf G} \cdot \mbox{\boldmath r} } 
 \,\, u \, K_1(u) 
 \over 
 \lambda_{ab}^2 G^2}, \eqnum{1a} \\
\nonumber \\
\mbox{where} \;\;\;\;\;
u^2 & = & 2 \, \xi_{ab}^2 G^2 (1+b^4)[1-2b(1-b)^2], \eqnum{1b}
\end{eqnarray} 
$B_0$ is the average magnetic field, 
{\bf G} are the reciprocal lattice vectors,
$b \! = \! B_{0}/B_{c_2}$,
$\xi_{ab}$ is the GL coherence length and $K_1(u)$ is a modified
Bessel function. We do not expect the conventional GL model 
to be valid deep in the superconducting state. However, this model
gives a very good fit to the measured field distribution---which is
all that is required
to generate the corresponding $J_s({\bf r})$ profile needed
to determine $r_0$. The summation in Eq.~(\ref{eq:Br}) is taken over all
reciprocal lattice vectors {\bf G} of a triangular vortex lattice.
This assumption is reasonable, because for field-cooled samples
the vortex lattice geometry at low $T$
is governed by the geometry of the lattice at the pinning 
temperature.
We have shown previously \cite{Sonier:94} that the vortex lattice 
in our YBa$_2$Cu$_3$O$_{6.95}$ crystals is strongly pinned at low $T$ and
remains so upon warming up to $T \! \approx \! 0.7~T_c$. 
The lattice in the detwinned crystal depins at similar high temperatures.
In a $d_{x^2-y^2}$-wave superconductor,
the vortex lattice is predicted to be nearly triangular at temperatures close 
to $T_c$ ({\it e.g.} see Ref.~\cite{Franz:97}). We note that
even if the lattice is not triangular, a good fit still yields the
appropriate $J_s ({\bf r})$ profile. 

All of the data were fit in the time domain with a
theoretical muon polarization function constructed from 
the field profile of Eq.~(\ref{eq:Br}). This was multiplied
by a Gaussian relaxation function $e^{-\sigma^2 t^2/2}$ to
account for any residual disorder in the vortex lattice 
and the contribution of the nuclear dipolar moments
to the internal field distribution.
The residual background signal was fit assuming a Gaussian broadened
distribution of fields. The Fourier transform (FT) of the muon precession
signal approximates the internal field distribution and resembles
the predicted asymmetric lineshape for an ordered lattice of vortices.
However, the FT suffers from noise and broadening effects associated
with the finite number of 
events and limited time range, so that the data must 
be analyzed in the time domain.


Figure~1(a) shows the first 1.5~$\mu$s of a typical muon precession signal
in YBa$_2$Cu$_3$O$_{6.95}$ displayed in a reference frame rotating at 3.3~MHz
below the Larmor precession frequency of a free muon.
The solid curve is a fit (actually performed over the first 6~$\mu$s) to the
theoretical polarization function assuming $\xi_{ab} \! = \! 54$~\AA, with
all other fitting parameters unconstrained. 
The difference between this fit and the measured spectrum is shown
in Fig.~1(b), compared to the same assuming $\xi_{ab} \! = \! 20$~\AA.
Note that the quality of the fit is most affected by a change in $\xi_{ab}$
at early time, where the amplitude of the signal originating from the vortex
lattice is largest.
The ratio of $\chi^2$ to the number of degrees of freedom (NDF) for fits
assuming different values of $\xi_{ab}$ is shown in Fig.~2(a) for two
different magnetic fields. Due to the high statistics of the measured
field distribution ({\it i.e.} typically consist of $2 \! \times \! 10^7$
muon decay events), $\chi^2$/NDF is greater than one. Figure~2(b) shows that
$\kappa \! = \! \lambda_{ab} / \xi_{ab}$ obtained from the same fits
also depends on $H$.

Figure~3(a) shows the temperature dependence of $r_0$ at $H \! = \! 0.5$~T
and 1.5~T in sample TW1. The error bars represent the 
statistical uncertainty in the fitted values of $\xi_{ab}$.
Both sets of data show a slight decrease in $r_0$ with decreasing
temperature which is essentially linear below 50K. The strength of the
term linear in $T$ is comparable to that previously reported
in YBa$_2$Cu$_3$O$_{6.60}$ \cite{Sonier:97b}, which indicates
that thermal vibrations are an unlikely source of the observed behaviour.
On the other hand, this linear term is
considerably weaker than in NbSe$_2$ \cite{Sonier:97,Miller:98}---a
result consistent with the prediction of Ref.~\cite{Hayashi:98}
that a vortex core containing only a few discrete
bound states will stop shrinking below a relatively high saturation
temperature. It should be noted, that the core is also predicted to shrink
with decreasing $T$ in a pure
$d_{x^2-y^2}$-wave superconductor \cite{Ichioka:96}, which may not contain
any localized quasiparticle states.   

Figure~3(b) shows the magnetic field dependence of $r_0$ extrapolated
to $T \!= \!0$. Included in this figure is data from all three samples.
Note that there is excellent agreement between the twinned and detwinned
crystals. In Ref.~\cite{Sonier:97b}
it was shown that detwinning a YBa$_2$Cu$_3$O$_{6.60}$ single crystal
had virtually no effect on the deduced values of $r_0$, even though
there was evidence for distortions in the vortex lattice
caused by twin boundary pinning.  
We find that the fitted parameter $\xi_{ab}$ changes as a function
of $T$ and $H$ in a manner similar to that of $r_0$, with
$\xi_{ab} \! \approx \! [0.828(23)r_0 \! + \! 0.72(1.24)]$~\AA. 
Thus at large $H$ where $r_0$ is small, we have 
$\xi_{ab} \! \approx \! r_0$.
It is important to note however, that unlike $r_0$,
the precise behaviour of $\xi_{ab}$ does depend on the theoretical model 
used for $B(\mbox{\boldmath r})$. Although the GL model is 
invalid here, our measurements in both NbSe$_2$ and 
YBa$_2$Cu$_3$O$_{7-\delta}$ qualitatively resemble
the $T$ and $H$ dependence of the core size predicted from 
a quasiclassical treatment of a vortex in both an $s$-wave 
and a $d_{x^2-y^2}$-wave superconductor \cite{Hayashi:98,Machida:98}.

Some of the data in Fig.~3(b) is replotted in Fig.~4, along with
the measurements of $r_0 (H)$ in YBa$_2$Cu$_3$O$_{6.60}$ 
(from Ref.~\cite{Sonier:97b}). As was the case in NbSe$_2$,
both $\lambda_{ab}$ and the ratio 
$\lambda_{ab}/r_0$ ($\sim \! \kappa$) at $T \! = \! 0$ are well described 
by relations linear in $H$, so that
\begin{equation}
r_0 (H) = \frac{\lambda_{ab}(H)}{\lambda_{ab}(H)/r_0(H)}
= r_0(0) \frac{[1+\beta H]}{[1+\gamma H]} \, . \eqnum{2} 
\label{eq:phenom}
\end{equation}
Recently the field dependence of $\lambda_{ab}$ has been attributed
to both nonlinear and nonlocal effects associated with a
$d_{x^2-y^2}$-wave order parameter \cite{Amin:98} and does
not strictly reflect the field dependence of the superfluid
density.
In sample TW1 which constitutes the most complete data set for   
YBa$_2$Cu$_3$O$_{6.95}$, 
$r_0 (0) \! = \! 120.7$~\AA, $\beta \! = \! 0.075$~T$^{-1}$ and
$\gamma \! = \! 1.82$~T$^{-1}$.
On the other hand, in the twinned YBa$_2$Cu$_3$O$_{6.60}$ sample
of Ref.~\cite{Sonier:97b} we find that
$r_0 (0) \! = \! 156.3$~\AA, $\beta \! = \! 0.094$~T$^{-1}$ and
$\gamma \! = \! 1.77$~T$^{-1}$. The solid curves in Fig.~4
represent Eq.~(\ref{eq:phenom}) with the corresponding values
of $r_0 (0)$, $\beta$ and $\gamma$. These extrapolations suggest that
the shrinking of the vortex cores saturates at large $H$.
Note that Eq.~(\ref{eq:phenom}) can be
written strictly in terms of the intervortex spacing $L$
(since $L \! \propto \! \sqrt{H}$).


Maggio-Aprile {\it et al.} \cite{Maggio:95} attributed the 
two peaks observed in the spectrum for tunneling 
into a vortex core of YBCO
at $T \! = \! 4.2$~K and $H \! = \! 6$~T,
to the lowest bound quasiparticle 
energy level $E_{1/2} \! = \! 5.5$~meV. 
Using the formula $E_{1/2} \! = \! 2 \mu \Delta_0^2/E_F$ from
Ref.~\cite{Caroli:64} and taking $\xi_{ab}$ at low $T$ to be equivalent
to the BCS coherence length $\xi_0 \! = \! \hbar v_f/ \pi \Delta_0$, gives 
$\xi_{ab} \! = \! (2 \hbar^2/m_e \pi^2 E_{1/2})^{1/2} \! \approx \! 17$~\AA.
This result ($\approx \! r_0$) is plotted in Fig.~4.
The agreement with the extrapolated curve from our $\mu$SR measurements
\cite{footnotehc2}
is striking and raises the possibility that the vortex cores in
YBa$_2$Cu$_3$O$_{6.95}$ are
conventional-like ({\it i.e.} they contain localized quasiparticle
states). Several theoretical studies ({\it e.g.} \cite{Franz:98})
have suggested that bound states may arise from a second component in 
the order parameter, induced by spatial variations in the
$d_{x^2-y^2}$-wave component in the vicinity of a vortex core.
If this is the case, the core expansion will
be accompanied by the formation of numerous bound states, which should
be detectable by STM. Thus far, STM has not been used to probe
the vortex structure at low magnetic fields in the HTSs.

In conclusion, we have observed a large increase in the size of the
vortex cores in YBa$_2$Cu$_3$O$_{6.95}$ at low magnetic fields,
similar to the behaviour reported in the conventional superconductor
NbSe$_2$. The expansion of the cores appears to be a general property
of superconductors in the vortex state. The agreement with
STM measurements based on a conventional treatment of the
vortex core in YBa$_2$Cu$_3$O$_{6.95}$, supports the existence
of localized states in the core.

We thank K.~Machida for helpful discussions.
This work was supported by NSERC
and DOE grant DE-FG05-88ER45353.
    
\newpage 
\begin{center} 
FIGURE CAPTIONS 
\end{center} 
 
Figure 1. (a) The muon precession signal in YBa$_2$Cu$_3$O$_{6.95}$ after
field cooling to $T \! = \! 5.8$~K in a magnetic field $H \! = \! 0.498$~T.
The solid line is a fit described in the text. 
(b) Difference between the measured
precession signal and the theoretical polarization function [in (a)] assuming
$\xi_{ab} \! = \! 54$~\AA~ (solid circles) and 
$\xi_{ab} \! = \! 20$~\AA~ (open squares). For visual clarity, we have doubled
the bin size used in (a) and not shown the error bars ($\approx \! \pm0.003$).\\

Figure 2. (a) The ratio $\chi^2$/NDF as a function of (a) $\xi_{ab}$
and (b) $\kappa$ (from the same fits) for $H \! = \! 0.498$~T (stars, 
NDF=1148) and $H \! = \! 1.952$~T (circles, NDF=1196) at $T \! = \! 5.8$~K.\\ 

Figure 3. (a) The temperature dependence of $r_0$ at 0.5~T (open circles)
and 1.5~T (solid circles) in twinned YBa$_2$Cu$_3$O$_{6.95}$ (TW1).
(b) The magnetic field dependence of $r_0$ extrapolated to $T \! = \! 0$.
The twinned crystals TW1 and TW2 are shown as open circles and squares,
respectively, whereas the detwinned crystal is denoted by solid triangles.\\

Figure 4. The magnetic field dependence of $r_0$ in YBa$_2$Cu$_3$O$_{6.95}$
(solid circles) and YBa$_2$Cu$_3$O$_{6.60}$ (open triangles) extrapolated 
to $T \! = \! 0$. The solid square at $H \! = \! 6$~T is the value of
$\xi_{ab} \! \approx \! r_0$ deduced from the STM experiment \cite{Maggio:95}
on twinned YBa$_2$Cu$_3$O$_{7-\delta}$ at $T \! = \! 4.2$~K. 
The solid curves are explained in the text.

 
\end{document}